\documentclass[prd,twocolumn,amsfonts,showpacs]{revtex4}
\newcommand{\beq}{\begin{equation}}
\newcommand{\eeq}{\end{equation}}
\newcommand{\beqs}{\begin{eqnarray}}
\newcommand{\eeqs}{\end{eqnarray}}
\newcommand{\lsim}{\mathrel{\raisebox{-
.6ex}{$\stackrel{\textstyle<}{\sim}$}}}
\newcommand{\gsim}{\mathrel{\raisebox{-
.6ex}{$\stackrel{\textstyle>}{\sim}$}}}

\begin{document}

\title{Upper Limits on a Possible Gluon Mass} 

\author{Shmuel Nussinov$^{a,b}$}

\author{Robert Shrock$^c$}

\affiliation{(a)
School of Physics and Astronomy, Tel Aviv University, Tel Aviv, Israel}

\affiliation{(b)
Schmid College of Science, Chapman University, Orange, CA 92866} 

\affiliation{(c)
C.N. Yang Institute for Theoretical Physics, Stony Brook University,
Stony Brook, NY 11794}

\begin{abstract}

We analyze upper limits on a possible gluon mass, $m_g$. We first discuss
various ways to modify quantum chromodynamics to include $m_g \ne 0$, including
a bare mass, a Higgs mechanism, and dynamical breaking of color SU(3)$_c$. From
an examination of experimental data, we infer an upper limit $m_g < O(1)$ MeV.
We discuss subtleties in interpreting gluon mass limits in view of the fact
that at scales below $\Lambda_{QCD}$, quantum chromodynamics is strongly
coupled, perturbation theory is not reliable, and the physics is not accurately
described in terms of the Lagrangian degrees of freedom, including gluons.  We
also point out a fundamental difference in the behavior of quantum
chromodynamics with a nonzero gluon mass and a weakly coupled gauge theory with
a gauge boson mass.

\end{abstract}


\maketitle

\section{Introduction} 

In quantum chromodynamics (QCD), a mass term for the gluon in Lagrangian is
forbidden by the color SU(3)$_c$ gauge invariance.  Experimental data are
consistent with the inference that the gluon mass $m_g$ is zero. But it is of
fundamental importance to inquire how stringent the experimental upper limits
are on a gluon mass and what the physical consequences of such a mass would be.
Considerable theoretical interest in this question was generated, starting in
the late 1970's \cite{dgj78}-\cite{gr82}, by a report of evidence for free
quarks, but later experiments did not confirm this report (some reviews are
\cite{qrev,pdg}). The literature on gluon mass limits is unsettled; published
upper bounds on $m_g$ range over ten orders of magnitude, from values of a few
MeV \cite{pdg} to $m_g < 2 \times 10^{-10}$ MeV \cite{yndurain}.  It is
obviously important to clarify this question, and we address it here.  As will
be evident from our discussion, the question is interesting partly because it
touches on some deeper conceptual issues, such as (i) how one can try to
construct a modification of QCD with a gluon mass small compared with the
confinement scale of $\sim (1 \ {\rm fm})^{-1}$; (ii) the question of the
extent to which one can get information on the Lagrangian fields in a confining
or quasi-confining theory; and (iii) the related quantum mechanical issue
pertaining to the accuracy with which one can measure properties of spatially
confined particles.

\section{Types of Modifications of QCD to Include a Gluon Mass}

\subsection{Bare Mass} 

It is first necessary to specify which type of modified QCD with nonzero gluon
mass or masses one considers.  There are several possible approaches to this.
First, one can consider a modification of QCD in which the Lagrangian ${\cal
L}_{QCD}$ contains a bare gluon mass term
\beq
{\cal L}_{QCD,m_g} = -\frac{m_g^2}{2} \sum_a A^a_\mu A^{a \ \mu} \ , 
\label{mg}
\eeq
where $a$ is the color index. Here, $m_g$ is a hard mass \cite{dynmass}, which
is present for arbitrarily weak QCD running coupling, $\alpha_s(\mu) \equiv
g_s(\mu)^2/(4\pi)$. (The scale $\mu$ will often be left implicit in the
notation.)  The mass term (\ref{mg}) explicitly breaks the SU(3)$_c$ gauge
invariance to a global SU(3)$_c$ symmetry.  One could also consider a more
general bare gluon mass term $-(1/2) \sum_a m_{g,a}^2 A^a_\mu A^{a \ \mu}$, but
the term in Eq. (\ref{mg}) will be sufficient for our purposes here.  

Formally, the inclusion of the gluon mass (\ref{mg}) in QCD this renders the
theory perturbatively non-renormalizable.  Thus, for example, if perturbative
methods were applicable and one were to compute amplitudes for longitudinally
polarized gluon-gluon scattering to multigluon final states, these would have
partial wave amplitudes that would involve powers of $s/m_g^2$ and hence would
violate perturbative unitarity when $\sqrt{s}$ exceeds a value of order $m_g$.
However, as discussed below, experimental data constrain $m_g$ to be less than
a few MeV, considerably less than the scale, $\Lambda_{QCD} \simeq 300$ MeV at
which $\alpha_s$ grows to O(1) and QCD exhibits the property of confinement or
quasi-confinement.  Here, we use the term ``quasi-confinement'' to mean that
free color-nonsinglet states have masses much larger than $\Lambda_{QCD}$ and
hence are integrated out of the modified QCD, defined as a low-energy effective
field theory.  An important point is that in the mass region well below
$\Lambda_{QCD}$, one cannot use perturbation theory.  One consequence of this
is that one cannot draw firm conclusions from the apparent violation of
perturbative unitarity in the above-mentioned partial wave amplitudes for
$\sqrt{s} \gsim m_g$.  Another is that although one can formally insert a
nonzero value of $m_g$ in ${\cal L}_{QCD,m_g}$, the physical meaning of this is
not completely clear, because one does not perform actual physical measurements
that are sensitive to this value if it is much less than $\Lambda_{QCD}$.

\subsection{Higgs Mechanism} 

  A second approach to modifying QCD to produce a gluon mass is to try to use a
Higgs mechanism, with a Higgs potential arranged so as to produce a vacuum
expectation value (VEV) of one or more color-nonsinglet Higgs fields,
spontaneously breaking SU(3)$_c$.  A scheme with three color triplets of Higgs
fields coupled in a manner such as to break the SU(3)$_c$ gauge symmetry to
global SU(3) was studied in Ref. \cite{dgj78}.  A second Higgs scheme is based
on the observation that the structure of the baryon wavefunction can be
retained if the breaking preserves an SO(3) subgroup of SU(3)$_c$, such that
the quarks transform as the vector representation, $\vec{q}$, of this SO(3);
this involves the equivalence of the wavefunction $\epsilon_{abc}q^a q'^b
q''^c$ in SU(3)$_c$ form with ${\vec q} \cdot ({\vec q} \, ' \times {\vec q} \,
'')$ in SO(3) form \cite{sgs81,qq}.  In this scheme the five gluons in the
coset space ${\rm SU}(3)_c/{\rm SO}(3)$ gain masses, while the three gluons
corresponding to the generators of SO(3) remain massless.  These three massless
gluons, $\vec g$, would also naturally form SO(3)-singlet bound states, ${\vec
g} \cdot {\vec g}$.  For this scheme, one must thus use a Higgs field that
contains a component transforming as a singlet under the SO(3) subgroup of
SU(3)$_c$.  The lowest-dimensional representation that has this property is the
27-dimensional representation of SU(3)$_c$ \cite{sgs81}.  A third type of Higgs
model is simpler in that it only uses two Higgs fields, both transforming as
fundamental representations of SU(3)$_c$. This breaks SU(3)$_c$ in two stages,
first to an SU(2) subgroup, and then completely, leading to two different
scales of masses for gluons, which may be comparable.

A Higgs mechanism for breaking SU(3)$_c$ and giving gluons masses has the
appeal that it preserves the renormalizability of the theory.  If one could
analyze the model perturbatively, the gluon mass(es) would be $\sim g_s|v|$,
where $v$ denotes a generic VEV of the colored Higgs field(s).  To illustrate
this, we may consider a very simple case with just one electroweak-singlet
Higgs field $\phi$ transforming as a fundamental representation of SU(3)$_c$,
with potential
\beq
V = \mu^2 \phi^\dagger \phi + \lambda (\phi^\dagger \phi)^2 
\label{v}
\eeq
where $\mu^2 < 0$. If one were able to use perturbation theory reliably here,
then the minimization of the potential $V$ would lead to a nonzero VEV for
$\phi$ given by $v \propto \sqrt{-\mu^2/\lambda}$. If, indeed, one were able to
do this, then, without loss of generality, one could choose the basis for
SU(3)$_c$ generators such that $\langle \phi \rangle_0 = (0,0,v)^T$.  This
would break SU(3)$_c$ to the SU(2)$_c$ subgroup generated by $T_a$, $a=1,2,3$
\cite{higgs27}. The five gluons in the coset ${\rm SU}_c(3)/{\rm SU}_c(2)$
corresponding to the generators $T_a$, $a=4,..,8$, would pick up masses $m_g
\sim g_s |v|$

However, there is an important difference between the attempt to use a Higgs
mechanism to break SU(3)$_c$ and the use of this type of mechanism to break
electroweak symmetry in the Standard Model (SM).  Given that, as discussed
further below, $m_g$ must be smaller than a few MeV, considerably below
$\Lambda_{QCD}$, the color-nonsinglet Higgs fields interact strongly, and one
cannot use perturbation theory to analyze their behavior.  In particular, one
cannot reliably conclude that setting $\mu^2$ to a negative value of magnitude
small compared with $\Lambda_{QCD}^2$ would actually lead to a nonzero VEV for
$\phi$ \cite{g80}.  This problem, by itself, is sufficiently severe to motivate
one to consider a different renormalizable mechanism for giving gluons a mass.

Secondly, although one cannot use perturbation theory reliably to calculate the
masses of residual Higgs fields, and hence they might be larger than the
perturbative expressions $m_H \sim \sqrt{\lambda}|v|$, they could well be
sufficiently light as to be excluded by experimental limits.  To obtain
properties of strongly coupled systems of gauge, Higgs, and fermion fields
requires a fully nonperturbative calculational method, and a lattice field
theory formulation can provide this. For a lattice theory with a Higgs field
transforming according to the fundamental representation of the gauge group the
confinement and Higgs phases are analytically connected in the absence of
fermions \cite{fs}, but are separated by a phase boundary when fermions are
included \cite{ls1}. Since the lattice formulation maintains exact local gauge
invariance, a Higgs VEV as conventionally defined in the continuum vanishes
identically; instead, one measures various gauge-invariant quantities, such as
the bilinear fermion condensate and fermion and Higgs masses \cite{ghfrev}.
One of the issues that these nonperturbative lattice studies confronted was the
question of where to take the continuum limit in the space of relevant lattice
parameters and the fact that some portions of lattice phase boundaries were
first-order, with finite correlation lengths, instead of second-order
transitions, with the infinite correlation length that is necessary to
construct a continuum limit free of lattice artifacts.  Notwithstanding this
complication, however, these lattice studies tended to find ratios of Higgs to
gauge boson masses which did not differ strongly from unity. One could thus use
the results from fully nonperturbative studies to support the concern that in a
Higgs picture the spectrum would contain bound states involving the Higgs
fields (with themselves or quarks) that are not seen experimentally,
disfavoring the Higgs approach to trying to produce nonzero gluon masses.

There are also several other problems with a color-nonsinglet Higgs mechanism.
The addition of such Higgs fields to QCD reduces the renormalization-group 
running of $\alpha_s(\mu)$ as a function of energy scale $\mu$.
The increase of $\alpha_s(\mu)$ as the energy scale decreases from $\mu=m_Z$
down to the scale of $b \bar b$ quarkonium states is consistent with $N_f=5$
dynamical quarks, and the further evolution down to the scale of $c \bar c$
quarkonium states is consistent with $N_f=4$ dynamical quarks
\cite{bethke,pdg}.  This agreement would be upset if one added too many
additional light color-nonsinglet Higgs to the theory.  Equally if not more
problematic is the fact that the quartic Higgs coupling is not asymptotically
free and hence grows as the energy scale increases, undermining the asymptotic
freedom of QCD and leading to a possible Landau singularity.  Furthermore, the
parameter $\mu^2$ is quadratically sensitive to ultraviolet physics; i.e.,
there is a hierarchy problem.  Because of all of these problems, it not clear
whether one could, in fact, use a Higgs mechanism to break the SU(3)$_c$ gauge
symmetry (either completely or to a nontrivial subgroup gauge symmetry) and
obtain the values of gluon masses of a few MeV.  This motivates one to consider
alternatives.

\subsection{Dynamical Breaking of SU(3)$_c$ and Generation of Gluon Mass}

There is a third way that one might try to break color SU(3)$_c$ which, to our
knowledge, has not received much attention in the literature, namely via the
formation of a color-nonsinglet bilinear fermion condensate produced by another
strongly coupled gauge interaction.  We will investigate this possibility here
using a simple model but will show that this model also has problems.  Let us
consider an extension of the Standard Model gauge group $G_{SM} = {\rm SU}(3)_c
\times {\rm SU}(2)_L \times {\rm U}(1)_Y$ (where $Y$ is the weak hypercharge)
in which we adjoin another gauge group SU(2)$_{mc}$, where $mc$ stands for
metacolor (not to be confused with technicolor).  Thus, the full gauge group
that is operative at scales above a few GeV is taken to be
\beq
G_{SM'} = {\rm SU}(2)_{mc} \times G_{SM} \ . 
\label{gsmprime}
\eeq
In addition to the usual SM fermions, we add left and right-handed 
electroweak-singlet chiral fermions (indicated with $L,R$) transforming as 
\beq
\zeta^{a \alpha}_L \ : \quad (2,3,1)_0 
\label{zeta}
\eeq
\beq
\eta^\alpha_L \ : \quad (2,1,1)_0
\label{eta}
\eeq
and
\beq
\chi^a_{p,R} \ : \quad 2(1,3,1)_0  \quad {\rm for} \ \ p=1,2 \ , 
\label{chi}
\eeq
where $a$ and $\alpha$ denote SU(3)$_c$ and SU(2)$_{mc}$ gauge indices,
respectively, the numbers in parentheses denote the dimensionalities of the
representations of ${\rm SU}(2)_{mc} \times {\rm SU}(3)_c \times {\rm
SU}(2)_L$, the subscript is the value of $Y$, and the set includes two copies
of the $(1,3,1)_0$ field labelled with a copy number $p=1,2$.  

By analogy with quarks, we assign baryon number $B=1/N_c=1/3$ to the
color-triplet fermions $\zeta^{a \alpha}_L$ and $\chi^a_{p,R}$.  In the
Lagrangian describing the high-scale physics, these fermions are taken to have
mass terms whose coefficients are zero or at least negligibly small compared
with $\Lambda_{QCD}$.  The color-singlet fermion $\eta^\alpha_L$ is included so
that there are an even number of left-handed SU(2)$_{mc}$ doublets, as is
required to avoid a global Witten anomaly associated with the homotopy group
$\pi_4({\rm SU}(2))={\mathbb Z}_2$.  The SU(2)$_{mc}$ gauge sector thus
contains four left-handed chiral fermions, or equivalently, two Dirac fermions,
transforming as metacolor doublets.  The resultant theory, consisting of these
fermions plus those of the Standard Model, is free of anomalies in all gauged
currents. The SU(3)$_c$ interaction remains vectorial and asymptotically free.
Since this model involves the introduction of two additional light flavors of
color-triplet Dirac fermions to QCD, it reduces the agreement with $N_f=5$
quarks that characterizes the measured dependence of
$\alpha_s(\mu)$ on the scale $\mu$ between $\mu=m_Z$ and $\mu \simeq 10$ GeV,
the scale characterizing the $b \bar b$ $\Upsilon$ states. But we will show
next that the model has even more serious problems.

Since the SU(2)$_{mc}$ gauge interaction is asymptotically free, as the
reference energy scale $\mu$ decreases from large values, its coupling
$\alpha_{mc}(\mu) \equiv g_{mc}(\mu)^2/(4 \pi)$ increases.  Let us first
consider the case where the value of $\alpha_{mc}(\mu)$ is sufficiently large
at a high scale $\mu >> \Lambda_{QCD}$ so that this coupling grows to values of
order unity at a scale $\Lambda_{mc} > \Lambda_{QCD}$. (The reason for this
assumption will be explained below.)  A study of the Dyson-Schwinger equation
for the fermion propagator in an asymptotically free vectorial SU($N$) gauge
theory (at zero temperature) with $N_f$ copies of massless fermions
transforming according to the fundamental representation of this gauge group
suggests that if $N_f < N_{f,cr}$, then as the theory evolves into the
infrared, it produces a bilinear fermion condensate that spontaneously breaks
the global chiral symmetry, whereas if $N_f > N_{f,cr}$, such a condensate is
not formed and instead the chiral symmetry remains exact, where $N_{f,cr}$ is a
certain critical number \cite{nfcr}.  For the case $N=2$ relevant here, a
solution of the Dyson-Schwinger equation in the one-gluon exchange
approximation yields the value $N_{f,cr} \simeq 8$.  Since this is well above
the number $N_f=2$ that we have in the SU(2)$_{mc}$ model, we can confidently
infer, given our assumption that, as the scale $\mu$ decreases, the metacolor
coupling gets large before the color coupling does, that the SU(2)$_{mc}$
interaction produces bilinear fermion condensates.  The most attractive channel
for these is $2 \times 2 \to 1$, where the numbers refer to the
dimensionalities of fermion SU(2)$_{mc}$ representations. These include a
condensate $\langle \epsilon_{\alpha\beta} \, \zeta^{a \alpha \ T}_L C \,
\zeta^{b \beta}_L \rangle$, where $\epsilon_{\alpha\beta}$ is the antisymmetric
tensor density for SU(2)$_{mc}$. This is automatically antisymmetrized in the
color indices $a,b$ and hence is proportional to
\beq
\langle \epsilon_{abc} \epsilon_{\alpha\beta} \, \zeta^{a \alpha \ T}_L C \, 
\zeta^{b \beta}_L \rangle
\label{zetazeta}
\eeq
where $\epsilon_{abc}$ is the antisymmetric tensor density for SU(3)$_c$.  The
condensate (\ref{zetazeta}) transforms as a color $\bar 3$ 
and hence dynamically breaks SU(3)$_c$ to an SU(2) subgroup.  A
second condensate formed by the SU(2)$_{mc}$ interaction is
\beq
\langle \epsilon_{\alpha\beta} \zeta^{a \alpha \ T}_L C \eta^{\beta}_L \rangle 
\ . 
\label{zetaeta}
\eeq
This transforms as a \underline{3} under SU(3)$_c$ and hence also breaks it to
an SU(2) subgroup.  One can use vacuum alignment arguments to infer that these
SU(2) subgroups are identical.  Then, without loss of generality, one may
choose the index $c=3$ in the condensate (\ref{zetazeta}) and $a=3$ in the
condensate (\ref{zetaeta}), so that the residual SU(2) subgroup of SU(3)$_c$
preserved by these condensates, which we will denote as SU(2)$_c$, is the one
for which the corresponding Lie algebra is generated by $T_a$ with $a=1,2,3$.
The five gluons in the coset ${\rm SU}(3)_c/{\rm SU}(2)_c$, i.e., those
corresopnding to $T_a$, $4 \le a \le 8$, gain masses of order
\beq
m_g \sim g_s(\Lambda_{mc}) \, \Lambda_{mc} \sim \Lambda_{mc} \ . 
\label{mcmg}
\eeq
The fermions involved in these condensates, $\zeta^{a \alpha}_L$ with $a=1,2,3$
and $\eta^\alpha_L$ for $\alpha=1,2$, also gain dynamical masses of order
$\Lambda_{mc}$,

This, then, is a renormalizable, dynamical way to break SU(3)$_c$ (to
SU(2)$_c$).  In contrast to a Higgs mechanism, it is technically natural and
does not suffer from any hierarchy problem.  This model shows that the property
that a gauge symmetry is vectorial is not sufficient, in itself, to ensure that
it remains unbroken.  Indeed, since the weak isospin SU(2)$_L$ gauge
interaction is asymptotically free, if it had not been broken at the
electroweak scale but instead had been able to grow in strength to a sufficient
level, it would have broken SU(3)$_c$ to SU(2)$_c$ in a somewhat analogous
manner \cite{smr}.  One could presumably add additional fields and/or
interactions to this metacolor model so that SU(3)$_c$ would be broken
completely.  However, although this dynamical approach avoids some of the
problems with the other approaches to breaking SU(3)$_c$ that we have described
above, the gluon masses that it produces, given in Eq. (\ref{mcmg}), are too
large to be allowed by experiment.  This is clear from an example.  Consider,
say, the value $\Lambda_{mc} = 10$ GeV.  Experimental data determine
$\alpha_s(10 \ {\rm GeV}) = 0.18$ (see, e.g., Fig. 6 of Ref. \cite{bethke}),
i.e., $g_s(10 \ {\rm GeV}) = 1.5$. Then from Eq. (\ref{mcmg}), we would obtain
$m_g \sim 15$ GeV, which is much too large to agree with experiment.  A second
illustrative choice is $\Lambda_{mc} = 1$ GeV. For this choice, one has
$\alpha_s(1 \ {\rm GeV}) \simeq 0.5$ \cite{bethke}; from Eq. (\ref{mcmg}) one
obtains $m_g \sim 2.5$ GeV, which is again much too large.  One cannot improve
this situation by selecting initial conditions for $\alpha_{mc}(\mu)$ at a high
scale $\mu$ so that $\Lambda_{mc}$ is smaller than $\Lambda_{QCD}$, because if
the SU(3)$_c$ interaction becomes strongly coupled, with $\alpha_s \sim O(1)$,
at a scale where the SU(2)$_{mc}$ interaction is still weakly coupled, then
among the bilinear fermion condensates produced by the QCD interaction, in
addition to $\langle \bar q_L q_R \rangle + h.c.$ for $q=u,d,s$, there would be
\beq
\langle \bar \zeta_{a, \alpha,L} \, \chi^a_{p,R} \rangle \ , \quad p=1,2 \ , 
\label{zetachi}
\eeq
which would break SU(2)$_{mc}$.  This is analogous to the fact that the QCD
quark condensates $\langle \bar q_L q_R \rangle + h.c.$ break SU(2)$_L$ (which,
however, is already broken at the much higher scale 250 GeV).  
Since the SU(2)$_{mc}$ symmetry would not be active in the low-energy effective
theory applicable at energy scales below $\Lambda_{QCD}$, its coupling 
$\alpha_{mc}$ would be frozen at this scale and hence would
not become large enough to break color.  Thus, although this model for 
dynamically breaking SU(3)$_c$ is renormalizable and does not have a hierarchy
problem, it is excluded by the fact that it would yield excessively large
values for the gluon masses.  It also would have the problem that it would
predict new hadronic states at experimentally accessible masses, and these have
not been observed.

One could also consider other mechanisms such as attempting to formulate QCD in
five or more dimensions and choosing boundary conditions in the
higher-dimensional space that break SU(3)$_c$ and give rise to a gluon mass in
the usual $(3+1)$-dimensional Minkowski space.  The fact that the
higher-dimensional theory is not renormalizable leads to a number of
complications, and we do not pursue this direction here.

Our analysis of various approaches to producing a gluon mass that is small
compared with $\Lambda_{QCD}$ has thus shown the difficulties that one
encounters with each of these approaches.  Although our analysis is not
exhaustive, it does show how challenging it is to construct a self-consistent
calculable model that could explain a gluon mass that is small compared with
the scale where QCD becomes strongly coupled and confines.  It is also worth
noting that the property $m_g=0$ is protected by the color gauge invariance,
and once this condition is removed, i.e., once one considers $m_g \ne 0$,
breaking SU(3)$_c$ gauge invariance, then there is no obvious symmetry that
could naturally keep $m_g$ small compared with other relevant scales, in
particular, $\Lambda_{QCD}$.

\section{An Upper Bound on $m_g$ from Experimental Data}

Here we step back from the construction of a model that could account naturally
for a small gluon mass and, in a more phenomenological framework, analyze how
large a value of $m_g$ might be allowed by experimental data.  To the extent
that we need a theoretical framework, we will use that given by the hard bare
mass term in Eq. (\ref{mg}), recognizing that it would require an ultraviolet
completion to answer the question of the origin of the gluon mass.  There are
many pieces of experimental evidence showing that $m_g$ must be considerably
smaller than $\Lambda_{QCD}$; the question is how much smaller. 

 By standard Bohr-Oppenheimer and effective field theory arguments, if a
particle has a mass $m$, then it does not play a dynamical role in the
low-energy effective theory that is operative at scales well below $m$.  It
follows quite generally that $m_g$ must be smaller than $\Lambda_{QCD}$ because
if it were not, then as the reference energy scale $\mu$ decreased, gluons
would be integrated out before $\mu$ decreased to $\Lambda_{QCD}$, and hence
$\alpha_s(\mu)$ would never grow to values of O(1).  A weakly coupled QCD with
a nonzero $m_g$ would not confine, so that there would be color-nonsinglet
physical states in nature.  (This would be analogous to the fact that since
weak-isospin SU(2)$_L$ is a broken gauge symmetry, it does not confine
neutrinos or charged leptons.)  But, in fact, there are no confirmed
observations of such states, in particular, free quarks, and there are quite
stringent upper limits on them, both from searches in matter and in collider
experiments \cite{qrev,pdg}.  Furthermore, analyses of the Dyson-Schwinger
equation for the quark propagator \cite{lane,politzer} have shown that if one
starts with a zero-mass quark, then, if $C_2(R)\alpha_s$ is greater than
roughly unity (where $C_2(R)$ is the quadratic Casimir invariant for the
representation $R$, equal to 4/3 for the fundamental representation of
SU(3)$_c$), this equation yields a solution with a nonzero value of the quark
mass.  This constitutes dynamical generation of a constituent quark mass, the
result of spontaneous chiral symmetry breaking in QCD.  That this is associated
with confinement can be understood by a simply physical argument \cite{casher}:
as a quark is headed outward from the center of a hadron and is reflected back
inward at the boundary, there is a flip of chirality, which amounts to the
presence of a $\bar q q$ term in the effective Lagrangian.  Since this
dynamical mass is the coefficient of $\bar q q$ in the effective QCD
Lagrangian, one may also associate this with the dynamical generation of a
condensate $\langle \bar q q \rangle$.  This spontaneous chiral symmetry
breaking gives rise to the approximate Nambu-Goldstone bosons of QCD, the
$\pi$, $K$, and $\eta$ mesons \cite{njl,kp}. However, if the gluons were
integrated out of the theory so that $\alpha_s$ never grew to values of O(1),
then this spontaneous chiral symmetry breaking would not occur.

In the presence of a nonzero gluon mass, even one much smaller than
$\Lambda_{QCD}$, the SU(3)$_c$ color gauge invariance is broken.  A consequence
of this is that color is, in fact, not completely confined.  This is obvious in
the conventional Higgs picture, given the inferred nonzero VEV of the
color-nonsinglet Higgs.  It also holds in the model presented above in which
there is dynamical breaking of SU(3)$_c$.  Consequently, an isolated quark or
gluon does not have infinite energy, and the Wilson-Polyakov line for such a
particle does not vanish identically.  If the limit $m_g/\Lambda_{QCD} \to 0$
is smooth, one expects that the mass of a free quark or gluon would diverge.
Thus, in a low-energy effective theory, these color-nonsinglet states would
disappear from the spectrum as $m_g \to 0$, and the resultant theory would
exhibit the property of being quasi-confining, in the sense that we defined
above.  Thus, for $m_g << \Lambda_{QCD}$, QCD, considered as a low-energy
effective theory, is quasi-confining.

It is worth noting that this expectation for the dependence of the mass of a
free quark or gluon on $m_g$ is borne out by a specific calculation within the
context of the MIT bag model.  Let us denote these states, dressed with their
cloud of glue, as $q_{dr}$ and $g_{dr}$.  The MIT bag model yields the masses
\cite{dgj78} 
\beqs
m_{q_{dr}} & = & \frac{\sigma}{m_g}
\left [ 1 + O \left ( (m_g/\sqrt{\sigma})^{1/3} \right ) \right ] \cr\cr
    & = & \frac{0.18 \ {\rm GeV}^2}{m_g} 
\label{mqmit}
\eeqs
and
\beq
m_{g_{dr}} = \frac{3}{2} m_{q_{dr}} \ , 
\label{mgmit}
\eeq
where
\beq
\sigma = \frac{1}{2 \pi \alpha'} = (420 \ {\rm MeV})^2 = 
0.90 \ {\rm GeV/fm}
\label{sigma}
\eeq
is the QCD string tension and $\alpha' = 0.9 \ {\rm GeV}^{-2}$ is the Regge
slope.  Because the string tension $\sigma \propto \Lambda_{QCD}^2$, these
masses can also be written in the form
\beq
m_{q_{dr}} = \frac{2}{3} m_{g_{dr}} = {\rm const.}\frac{\Lambda_{QCD}^2}{m_g}
\left [ 1 + O \left ( (m_g/\sqrt{\sigma})^{1/3} \right ) \right ] \ . 
\label{mqlambda}
\eeq
These estimates show how the limit $m_g/\Lambda_{QCD} \to 0$ can be
smooth, in the sense of low-energy effective field theory, since as $m_g \to
0$, the masses of a free quark or gluon diverge and they are integrated out of
the low-energy theory.  The physical, finite-mass states in this low-energy
theory are color-singlets.

Searches for free quarks in collider experiments depend on assumptions about
their electroweak transformation properties and decays \cite{pdg,cdfbprime};
current lower limits from collider searches on a quark of charge 2/3 or $-1/3$
vary between about 200 GeV and 340 GeV.  Taking the lower bound of 300 GeV as a
representative illustrative value and inserting this into Eq. (\ref{mqlambda}),
one obtains the nominal upper bound $m_g \lsim 0.5$ MeV.  Since there are
model-dependent aspects to the MIT bag model estimates of the masses of a free
quark or gluon, it is appropriate to allow a factor of a few to represent the
theoretical uncertainty, and also a similar factor to represent the effect of 
model-dependent assumptions in the limits obtained from experimental searches.
Including these, we infer that a reasonable upper bound on a possible gluon
mass is 
\beq
m_g < \ O(1) \ {\rm MeV} \ . 
\label{mglimit}
\eeq

It is also useful to estimate the dependence of the size of the $q_{dr}$ and
$g_{dr}$ states on $m_g$ for $m_g << \Lambda_{QCD}$, one sets the masses in
Eq. (\ref{mqlambda}) equal to the approximate volume $(4\pi/3)r^3$ times the
energy density, set by $\Lambda_{QCD}^4$, and hence obtains
\beqs
r & \sim & \left ( \frac{1}{m_g \Lambda_{QCD}^2} \right )^{1/3} \cr\cr
  & \sim & \frac{1}{\Lambda_{QCD}}\left (\frac{\Lambda_{QCD}}{m_g}\right)^{1/3} \ , 
\label{rmit}
\eeqs
i.e., $r \sim {\rm 1 \ fm}(\Lambda_{QCD}/m_g)^{1/3}$.  Hence, for $m_g$ small
compared with $\Lambda_{QCD}$, the sizes of deconfined, dressed quarks and
gluons would be substantially larger than the typical 1 fm size of a usual
hadron.

Another approach to the question of an upper limit on a gluon mass is to study
the effects of a nonzero $m_g$ on the static quark potential between a very
heavy quark $Q$ and antiquark $\bar Q$.  The short-distance part of this
potential for $r << \Lambda_{QCD}^{-1}$ has the Coulombic form
\beq
V_{Q \bar Q,sd}(r) = \frac{(4/3)\alpha_s(r)}{r} \ , 
\label{vqqbshort}
\eeq
and because short distances are equivalent to large $\mu$,
a nonzero $m_g$ that is small compared with $\Lambda_{QCD}$ would not affect
this significantly.  In regular QCD,
\beq
V_{Q \bar Q} = \sigma \, r \quad {\rm for} \ r \gsim \Lambda_{QCD}^{-1} \sim
1 \ {\rm fm} \ . 
\label{vqqbarlong}
\eeq
This linear growth in $V_{Q \bar Q}$ corresponds to the property that a
chromoelectric flux tube with energy per unit length $\sigma$ stretches between
the $Q$ and $\bar Q$. Making $m_g$ nonzero changes this so that for $r \gsim
m_g^{-1}$, $V_{Q \bar Q}(r)$ is damped by a factor $e^{-m_gr}$ and hence
decreases to zero for large $r$ rather than increasing without bound.  In turn,
this implies that $V_{Q \bar Q}(r)$ reaches a maximum at some value of $r \sim
m_{g}^{-1}$, where the force between the $Q$ and $\bar Q$, $\vec{F} = - \vec
\nabla V_{Q \bar Q}(r)$ vanishes. This is another indication that once $m_g$ is
nonzero, QCD no longer precisely confines, since a quark can tunnel through
this potential barrier.  If QCD only had very heavy quarks, then an analysis of
this static quark potential could provide a useful guide to an upper limit on
$m_g$.

However, real QCD has light quarks.  This has two effects: first, one cannot
use nonrelativistic quantities such as a potential energy associated with a $q
\bar q$ state reliably, because the physics is relativistic, and second, the
chromoelectric flux tube forming the string breaks in the process of
hadronization.  That is, when an actual $q \bar q$ pair is produced in a
reaction like $e^+e^- \to q \bar q$, as the $q$ and $\bar q$ separate to a
distance $r \sim \Lambda_{QCD}^{-1} \sim 1$ fm, and hence the energy in the
chromoelectric flux tube (string) is sufficient to produce hadronic final
states, such as $2\pi, \ 4\pi$, etc. it is energetically favorable for the
string to break with production of additional light $q \bar q$ pairs and
gluons, followed by hadronization.  The presence of a string extending to a few
fm can be interpreted as being responsible for short-lived hadronic resonances
lying on Regge trajectories up to masses of several GeV.  But the the string
(chromoelectric flux tube) does not stretch beyond a few fm;
instead, it is divided into smaller string bits as the $q \bar q$ pairs are
created.  For the relevant range of $m_g$ of a few MeV, recalling that $(1 \
{\rm MeV})^{-1} = 200$ fm, the string-breaking and $q \bar q$ pair creation and
resultant hadronization occur before the $e^{-m_gr}$ factor becomes
relevant. If, nevertheless, one were to attempt to apply a static quark
potential assuming no string breaking out to distances of order $m_g^{-1}$, one
would obtain apparently quite stringent apparent upper bounds on $m_g$
\cite{yndurain}.

The hadronization process can be modelled approximately via a non-Abelian
extension of the Schwinger mechanism \cite{cnn}.  Although this has not
been calculated for $m_g \ne 0$, a rough estimate of the effect of a gluon 
mass can be obtained from the result for $q \bar q$ production by the 
Schwinger mechanism \cite{cnn}, 
\beq
\frac{dW}{d^4x} \simeq \frac{\xi^2 }{4\pi^2} \sum_{n=1}^\infty \frac{1}{n^2}
\exp \left ( \frac{-n \pi m_q^2}{\xi} \right ) \ , 
\label{dwdv}
\eeq
where $\xi \equiv (g_s/2) {\cal E}$, with ${\cal E}$ serving as a measure of
the magnitude of the chromoelectric field in the flux tube (a general
expression in terms of quantities that are manifestly gauge-invariant and
Lorentz-invariant is given in \cite{ns}), and $m_q$ is the quark mass.  We
denote the area of a cross section of the flux tube by $A$. The string tension,
is given by $\sigma \sim ({\cal E}^2/2)A$, and Gauss' law implies that ${\cal
E}A=g/2$; combining these to eliminate $A$ and using the fact that
$\sigma=1/(2\pi \alpha')$, one obtains 
\beq
\xi = (g_s/2){\cal E} \simeq 2 \sigma = 0.35 \ {\rm GeV}^2  \ . 
\label{xivalue}
\eeq
It is plausible that the kinematic dependence of $dW/d^4x$ on $m_g$ would be 
somewhat similar to the dependence on $m_q$. We shall 
assume this and require that $dW/d^4x$ not change by more than a small
fractional amount $\epsilon$. For a rough bound, we retain just the first term
in the sum (\ref{dwdv}), which is the dominant term, and we require that the
fractional change in this term be less than $\epsilon$, i.e., 
$1-\exp(-\pi m_g^2/\xi) < \epsilon$.  This yields the upper limit 
\beq
m_g < \left [\frac{\xi}{\pi} \ln \left ( \frac{1}{1-\epsilon} \right ) \right
]^{1/2} 
\label{mgbound}
\eeq
With the above estimate for $\xi$ and the illustrative value 
$\epsilon = 0.01$, this yields the upper bound $m_g \lsim 35$ MeV, a 
somewhat less stringent bound than was obtained in (\ref{mglimit}).

There is currently no evidence for the proton decay or decays of neutrons that
are stably bound in nuclei, with typical partial lifetime limits $\tau/B >
10^{33} - 10^{34}$ yrs, where $B$ denotes the branching ratio for the given
mode \cite{pdg}.  Different types of SU(3)$_c$ breaking and gluon mass
generation yield different predictions for how this would change.  Thus, the
binding of protons would disappear if all gluons got masses of order
$\Lambda_{QCD}$, or if SU(3)$_c$ were broken to an SU(2) subgroup, but would
remain if SU(3)$_c$ were broken to a gauged SO(3) subgroup.  Thus, although in
principle one could use limits on proton and bound neutron instability to
constrain $m_g$, the results would depend strongly on the assumed type of
SU(3)$_c$ color breaking and gluon mass generation.  Let us consider the case
where SU(3)$_c$ is either broken completely or broken to an SU(2), rather than
SO(3) subgroup.  Then a proton could decay via a tunnelling process in which a
quark tunnelled out.  However, this tunnelling process would be very different
from the tunnelling mechanism responsible for $\alpha$-decays of heavy nuclei.
In the $\alpha$ decays of heavy nuclei, the emitted $\alpha$ particle has
essentially the same mass that it has inside the parent nucleus.  In contrast,
for the relevant range of $m_g \lsim O(1)$ MeV given by (\ref{mgbound}), a $u$
or $d$ quark with a current-quark mass of about 5 or 10 MeV and a constituent
mass of about 330 MeV inside a proton would have a mass of order hundreds of
GeV outside of the proton.  Clearly, not only would there be suppression of the
tunnelling process that might give rise to this emission of a quark, but also
it would be energetically forbidden.

Other effects of SU(3)$_c$ color breaking and nonzero gluon masses would occur
in the early universe.  Here, however, the temperature is finite rather than
zero, so that, strictly speaking, one would not be dealing with a
Lorentz-invariant gluon mass, but rather a gluonic screening mass.  For the
relevant range of allowed values of $m_g$ given by Eq. (\ref{mgbound}), which
are considerably below $\Lambda_{QCD}$, it follows that the finite-temperature
phase transition in the early universe would occur at a temperature $T_c \simeq
200$ MeV where SU(3)$_c$-breaking effects were negligible.  Hence, the
formation of free quarks in the early universe would mostly occur starting from
color-singlet states.  As in our discussion above, this formation process
depends on assumptions about how far the string between $q$ and $\bar q$
stretches before it breaks.  Owing to this and other model-dependent features
of the calculation, there is, for a given $m_g$, a wide range of possible
predictions for the resultant ratio in the number density of free quarks to
baryons, $n_q/n_B$.  For $m_g \sim$ few MeV, using results from
Ref. \cite{cnn}, Ref. \cite{kst81} found that $n_q/n_B$ could be in accord with
experimental bounds of order $10^{-22}$ \cite{kst81}. (See Ref. \cite{pdg} for
current upper limits on $n_q/n_B$.) We are in agreement with Ref. \cite{kst81}
but note that Ref. \cite{yndurain}, assuming considerably longer string
persistence lengths, claimed the much more stringent limit $m_g < 2 \times
10^{-10}$ MeV.

\section{On the Measurability of a Small Gluon Mass} 

The rough upper limit (\ref{mglimit}) shows that $m_g$ must be small compared
with $\Lambda_{QCD}$.  To what extent can one make this more precise?  In this
section, we address this question and stress that there is a basic problem that
one encounters in trying to do this.  Our starting point is the property that
QCD with $m_g=0$ (at zero temperature) confines. It may be recalled that in
addition to the experimental evidence, a convincing theoretical understanding
of this has come from the lattice gauge theory formulation of the theory.
Because the measure of the Euclidean QCD path integral on the lattice is
compact, one avoids inserting a Faddeev-Popov determinant in this measure and
maintains exact gauge invariance at all stages of the calculation.  One can
then rigorously define an order parameter for confinement, namely, the Wilson
loop.  The area-law behavior of the Wilson loop at strong bare coupling, i.e.,
small $\beta = 2N_c/g_0^2$, in conjunction with numerical simulations that
suggest that one can analytically continue from this limit to the continuum
limit at $\beta \to \infty$ constitute strong evidence that the continuum QCD
defined in this limit confines color.  The physical picture for this is the
chromoelectric flux tube that extends between an infinitely heavy, static quark
and antiquark, producing a static quark potential (\ref{vqqbarlong}), which
grows without bound as $r \to \infty$.  We have seen how, for $m_g <<
\Lambda_{QCD}$, although the modified QCD does not precisely confine, the
deconfined quarks and gluons have masses that are much larger than
$\Lambda_{QCD}$.  Hence, insofar as one deals with QCD as an effective
low-energy theory, the physical states in this theory with masses that are of
order $\Lambda_{QCD}$ or at least not many orders of magnitude larger than this
scale are color-singlets.  But this means that in this effective low-energy
theory, the physics is well described at realizable energies by a confining
theory and not by the Lagrangian fields, the quarks and gluons. This statement
becomes progressively more accurate as $m_g/\Lambda_{QCD}$ decreases toward
zero.  This suggests that it would be futile to try to set an upper limit on a
gluon mass that is many orders of magnitude smaller than $\Lambda_{QCD}$
because there is there is no well-defined gluon in the effective QCD theory
that is applicable in this energy range.  

Indeed, it follows that because gluons are quasi-confined, basic quantum
mechanics places a limit on how precisely one can probe for a nonzero but small
$m_g$ and prevents one from setting an upper limit on $m_g$ that is many orders
of magnitude less than $\Lambda_{QCD}$.  Since the color-singlet hadrons have
sizes of order $1/\Lambda_{QCD} \sim 1$ fm and the gluons are effectively
confined within a distance of this order, the Heisenberg uncertainty principle
dictates that one cannot, even in principle, measure the gluon momentum or
energy to a better accuracy than $\Delta |{\vec k}_g| \sim \Lambda_{QCD}$ and
$\Delta E_g \sim \Lambda_{QCD}$.  Hence, from such a measurement, one cannot,
in principle, distinguish between the case where $E_g = \sqrt{|{\vec
k}_g|^2+m_g^2}$ and the case where $E_g = |{\vec k}_g|$ for $m_g <<
\Lambda_{QCD}$.  Confinement in the effective QCD theory also implies a minimum
bound-state gluon momentum ${\vec k}_g$ of order $\Lambda_{QCD}$
\cite{lmax}. Indeed, the confined gluon propagator does not have a pole, and
hence the gluon does not have a well-defined mass.

 In principle, one might attempt to calculate glueball masses as a function of
$m_g$, then compute how their mixing with $q \bar q$ mesons to form mass
eigenstates changes as a result of varying $m_g$, and then compare the results
with experimental data to derive an upper bound on $m_g$.  For many
calculations of QCD properties, the lattice formulation is the appropriate
tool.  As noted above, the most natural formulation of lattice QCD maintains
exact local gauge invariance, and one would have to give up this advantage if
one were to try to use the lattice to study glueballs in the case of a nonzero
gluon mass, since this mass breaks the color gauge invariance.  Lattice QCD
calculations of glueball masses have been done for pure glue or glue with
quenched, but not light dynamical fermions \cite{gbmass}.  Ideally, one would
do this calculation with light dynamical fermions, compute the mixing matrix
that maps the (isosinglet) $J^{PC}=0^{++}$ states of $(|u \bar u\rangle + |d
\bar d\rangle)/\sqrt{2}$, $|s \bar s\rangle$, and $|{\rm glue} \rangle$ to the
$f_0$ mesons $f_0(1370)$, $f(1500)$, and $f(1710)$ (among others) and then
compare with experiment. However, there is no consensus what this mixing matrix
is experimentally, even for regular QCD with $m_g=0$ \cite{glueballs}.  
Another idea would be to try to look for some kinematic signature of a small
nonzero $m_g$ in hadron decays, similar to a test for quark masses in $D_s$
decays \cite{mn}, using helicity suppression arguments.  But the situation is
not analogous because of the presence of gluonic self-coupling and resultant $g
\to gg$ transitions.

\section{Contrast with Estimates of Quark Masses} 

It is of interest to contrast the situation concerning an upper bound on a
possible gluon mass with estimates of what are denoted the current- or ``hard''
masses of the light quarks $u$ and $d$ and the intermediate-mass quark $s$
\cite{hardmass}.  Here, these ``hard'' masses are to be distinguished from the
``soft'' constituent masses of order $\Lambda_{QCD}$ that are generated
dynamically for the light quarks by the formation of the quark condensates
$\langle \bar q q \rangle$ that spontaneously break chiral symmetry.  The key
point here is that in the limit in which one turns off color gauge
interactions, quarks still have weak and electromagnetic interactions, but
there are no gluons, since the gluons only enter as the gauge bosons of QCD.
The current or ``hard'' masses of the quarks are, indeed, defined as the masses
that these particles would have in the hypothetical limit in which QCD is
turned off \cite{hardmass}.  It has been challenging to determine the
current-quark masses $m_u$ and $m_d$ of the light quarks $u$ and $d$.  Two
further differences with respect to the gluons have enabled one to obtain
approximate values for these.  First, from the nucleon masses $m_p$
and $m_n$, one can infer that $m_u$ and $m_d$ differ only by a few MeV and that
$m_d > m_u$.  Second, because of spontaneous chiral symmetry breaking, one has
Gell-Mann-Oakes-Renner (GMOR) relations such as \cite{gmor}
\beq
m_\pi^2 = -\frac{(m_u + m_d)}{f_\pi^2} \langle \bar q q \rangle \ . 
\label{mpipsq}
\eeq
The measured values of $m_\pi$ and $f_\pi$, together with a determination of
$\langle \bar q q \rangle$ from, e.g., the lattice, then yield the value of
$m_u+m_d$.  From the corresponding GMOR relations for $m_{K^{+}}^2$ and
$m_{K^{0}}^2$, assuming flavor independence of $\langle \bar q q \rangle$ for
$q=u, \ d, \ s$, one can obtain approximate values for the ratios $m_d/m_u
\simeq 2$ and $m_s/m_d \simeq 20$ (e.g., \cite{gl}).  The fact that, even with
these methods and modern refinements \cite{gl}, there is still non-negligible
uncertainty in $m_u$ and $m_d$ shows the difficulty of extracting (hard) masses
of light confined particles.

\section{High-Energy Behavior of Cross Sections in QCD with $m_g \ne 0$} 

We elaborate here on an interesting point that we noted at the beginning of
this paper.  Let us consider a modified QCD theory with the nonzero $m_g$ in
Eq. (\ref{mg}) satisfying the bound (\ref{mgbound}). We treat this theory as an
effective field theory, valid up to some UV cutoff $\Lambda_{UV}$.  In order
for it to be a useful theory and to match experimental data, it is necessary
that $\Lambda_{UV} >> \Lambda_{QCD}$. This condition should hold if
$m_g/\Lambda_{QCD} << 1$.  A very interesting aspect of this construction is
its contrast with the Higgs mechanism in a weakly interacting theory, such as
the Standard Model.  Taking the limit of large Higgs mass in the SM, one
obtains an estimate of the highest energy to which the resultant theory
(denoted $SM'$) can be used as a perturbatively calculable effective field
theory, namely $\Lambda_{UV,SM'} \simeq 4\sqrt{\pi} \, v_{EW} =
8\sqrt{\pi}m_W/g = 1.7$ TeV \ \cite{lqt}, where $g$ is the weak SU(2)$_L$ gauge
coupling and $v_{EW} \equiv 2^{-1/4}G_F^{-1/2} = 246$ GeV.  For $\sqrt{s} \gsim
\Lambda_{UV,SM'}$, the $J=0$ partial wave amplitude for longitudinal vector
boson scattering violates perturbative unitarity, indicating the onset of
strongly coupled physics. However, the analogous procedure is not applicable in
our present case of QCD with $m_g \ne 0$, because for the relevant range given
by Eq. (\ref{mgbound}), longitudinal gluon-gluon scattering is not
perturbatively calculable, as a result of the strong coupling $\alpha_s \sim
O(1)$. The non-applicability of the perturbative partial wave amplitude
analysis to QCD with $m_g \lsim O(1)$ MeV is clear, because if one were able to
apply it, then, in terms of the color-nonsinglet Higgs VEV $|v|$, one would get
$\Lambda_{UV} \sim |v| \sim m_g/g_s$. But this result would not make physical
sense, since it would imply that, for example, QCD with $m_g = 1$ eV would
break down at a scale of order 1 eV.  In QCD with $m_g \ne 0$, the expression
for $\Lambda_{UV}$ has a form that is fundamentally different from the form
$\Lambda_{UV} \sim m_g/g_s$ that one would obtain for a perturbative theory.
The property that QCD with $m_g \ne 0$ matches onto the theory with $m_g=0$ as
$m_g \to 0$ implies that QCD with $m_g \ne 0$ should be a good effective theory
up to a scale $\Lambda_{UV}$ of the form
\beq
\Lambda_{UV} \sim \Lambda_{QCD} \left ( \frac{\Lambda_{QCD}}{m_g} \right )^\nu
\label{lambdau}
\eeq
where the exponent $\nu > 0$.  The strong-coupling nature of the theory in the
region of energies $\sqrt{s} \sim \Lambda_{QCD}$ makes it difficult to obtain a
precise value of the exponent $\nu$, but it is plausible that $\nu \sim O(1)$.
This value is in accord with the MIT bag model estimate of the masses of free
quarks and gluons in Eqs. (\ref{mqmit}) and (\ref{mgmit}).

\section{Contrast with Limits on a Photon Mass}

There is a striking contrast in the modest upper limit of a few MeV that one
can obtain for $m_g$ and the very stringent upper limit on the photon mass,
$m_\gamma \lsim 10^{-19}$ eV \cite{gn}. The fact that the upper bound on
$m_\gamma$ is so much smaller than the upper bound on $m_g$ can be traced to
the property that the photon is not confined, together with the property that
matter is electrically neutral on a macroscopic scale, and the ability to
observe electromagnetic fields, such as those associated with planetary dipole
fields and the solar wind, that have quite large spatial extent. Note that
conditions other than confinement could limit one's ability to set a bound on
the photon mass.  For example, consider the hypothetical situation in which one
were restricted to making observations in the interior of a metal, where,
instead of freely propagating photons, there are plasmons with plasma frequency
$\omega_p$ given by $\omega_p^2 = 4 \pi n e^2/m_e$ ($n=$ number density of
electrons).  Then one would only be able to obtain an upper bound on $m_\gamma$
that was a small fraction of $\omega_p$ (where $\hbar \omega_p \sim $ few eV in
typical metals). A similar comment would apply if one were restricted to making
observations in a medium where there is Debye screening.

\section{Conclusion} 

In this paper we have revisited the question of upper limits on a possible
gluon mass.  We have discussed various ways of modifying QCD to produce gluon
masses.  From an analysis of experimental constraints, we have concluded that a
reasonably robust upper bound is $m_g < O(1)$ MeV, given in
Eq. (\ref{mgbound}).  We have discussed some of the subtle conceptual issues
that one must confront in trying to set an upper bound on $m_g$ that would be
very far below the scale, $\Lambda_{QCD} \simeq 300$ MeV.  These include the
fact that in this mass range one cannot use perturbation theory reliably and
the physics is not accurately described in terms of the Lagrangian degrees of
freedom, including gluons.  Since the inapplicability of perturbation theory
makes it difficult to use a Higgs mechanism reliably to produce a small gluon
mass, we have explored how one might do this with a nonperturbative dynamical
mechanism and have shown how this attempt would yield excessively large values
of $m_g$.  We have shown how quasi-confinement in QCD with a small gluon mass,
in conjunction with the Heisenberg uncertainty principle, renders it difficult
to set an upper limit on $m_g$ that is very small compared with
$\Lambda_{QCD}$.  As part of our analysis, we have also shown that the
ultraviolet cutoff $\Lambda_{UV}$ on QCD with $m_g \ne 0$, considered as an
effective field theory, has a very different form from the ultraviolet cutoff
in the electroweak theory with a heavy Higgs.

\bigskip

Acknowledgments: The research of R.S. was partially supported by the grant
NSF-PHY-06-53342 (R.S.).

\vfill
\eject

\begin{thebibliography}{99}
\bibitem{qq}
%
It should be noted, however, that the SO(3) gauge symmetry would also allow
SO(3)-singlet but fractionally charged diquark bound states ${\vec q} \cdot
{\vec q} \, '$ and quark-gluon bound states ${\vec q} \cdot {\vec g}$, that are
forbidden by SU(3)$_c$ invariance.
\bibitem{higgs27} 
%
We note that it would be very difficult to construct a dynamical model in which
the condensates would break SU(3)$_c$ in the same manner as a Higgs field in
the $\underline{27}$ representation of SU(3)$_c$.  The reason for this is that
these condensates would form in the most attractive channel. A criterion for
the attractiveness of a channel involving fermions in representations $R_1$ and
$R_2$ of a given gauge group producing a condensate in the representation
$R_{cond.}$ is $\Delta C_2 = C_2(R_1) + C_2(R_2) - C_2(R_{cond.})$, where
$C_2(R)$ is the quadratic Casimir invariant for $R$. Thus, for a given $R_1$
and $R_2$, this favors the smallest possible $R_{cond.}$.  To get
$R_{cond.}=\underline{27}$ of SU(3)$_c$, one would have to start with $R_1$
and/or $R_2$ in very large representations, which would be disfavored 
by the resultant large effect on the QCD beta function.
\end{thebibliography}
\end{document}